\documentclass[12pt]{article}
\usepackage{a4wide,amssymb,cite}
\usepackage{color}
\newcommand{\be}{\begin{equation}}
\newcommand{\ee}{\end{equation}}
\newcommand{\bea}{\begin{eqnarray}}
\newcommand{\eea}{\end{eqnarray}}
\def\eq#1{eq.~(\ref{#1})}

\def\circa#1{\,\raise.3ex\hbox{$#1$\kern-.75em\lower1ex\hbox{$\sim$}}\,}
\def\mpl{M_{\rm Pl}}

\begin{document}

\begin{titlepage}

\rightline{CERN-PH-TH/2014-025}

\begin{centering}
\vspace{1cm}
{\large {\bf Starobinsky-like inflation from induced gravity}} \\

\vspace{1.5cm}

 {\bf Gian F. Giudice}$^{1,2}$ and {\bf Hyun Min Lee}$^{3}$ \\
\vspace{.5cm}

{\it $^1$CERN, Theory Division, CH-1211 Geneva 23, Switzerland} 
\\  \vspace{.5mm}
{\it $^2$Solvay Institute, Boulevard du Triomphe, 1050 Bruxelles, Belgium}
\\  \vspace{.5mm}
{\it $^3$Department of Physics, Chung-Ang University, Seoul 156-756, Korea}  \\

\vspace{.1in}

\end{centering}
\vspace{2cm}

\begin{abstract}
\noindent
We derive a general criterion that defines all single-field models leading to Starobinsky-like inflation and to universal predictions for the spectral index and tensor-to-scalar ratio, which are in agreement with Planck data. Out of all the theories that satisfy this criterion, we single out a special class of models with the interesting property of retaining perturbative unitarity up to the Planck scale. These models are based on induced gravity, with the Planck mass determined by the vacuum expectation value of the inflaton. 

\end{abstract}

\vspace{3cm}

\end{titlepage}

\section{Introduction}

The recent data from the Planck mission~\cite{planck} corroborate the case for a relatively simple universe, with nearly gaussian and adiabatic cosmological perturbations. These observations can be accounted for by the simplest form of inflation, driven by a single scalar field. However, the strong constraint on the tensor-to-scalar ratio $r$ from Planck ($r< 0.11$ at 95\% CL~\cite{planck}) disfavours single-field inflationary models with minimal coupling to gravity and polynomial potentials. On the other hand, Starobinsky-like models, based on an effective $R+R^2$ gravity~\cite{r2inflation} (for a recent review, see ref.~\cite{linde}), predict the following values of $r$ and of the spectral index $n_s$ as a function of the number of e-folds $N$:
\be
r=\frac{12}{N^2}\, ,~~~n_s=1-\frac{2}{N}\, .
\label{star}
\ee
As Planck $+$ WMAP polarisation data \cite{planck} indicate $n_s=0.9603\pm 0.0073$ at $68\%$ CL, the prediction of the Starobinsky-like models is successful at 95\% CL for $40\circa< N\circa< 80$.

In this paper we give a general criterion that
fully characterises the class of single-field inflationary models leading to the Starobinsky-like relations in \eq{star}. 
Particular cases of Starobinsky-like theories are the ``universal attractor" models~\cite{attractors}, Higgs inflation~\cite{higgsinf} (for a recent review, see ref.~\cite{hrev}) and S-inflation~\cite{sinflation} which, as  shown in ref.~\cite{riotto}, are all equivalent to an effective $R+R^2$ gravity as far as inflationary dynamics is concerned. Moreover, Starobinsky models have similarities to extra-dimension moduli~\cite{extrad1,extrad2} and the holographic dual in five dimensions~\cite{kiritsis}. 

It is also well known (see refs.~\cite{unitarybound} and \cite{riotto}) that the Starobinsky-like models generally violate perturbative unitarity at an energy scale much lower than the Planck mass. In many cases, the scale of unitarity violation is lower than the inflationary scale, casting serious doubts on the reliability of the calculation (and the beneficial effects from the field dependence of the cut-off scale~\cite{bmass} are usually not sufficient to resolve the problem). Moreover, losing perturbative unitarity has another drawback, which is quite independent of the issue about calculability of inflationary dynamics. The lack of new-physics discoveries at the LHC has fuelled speculations that the Standard Model (SM), or one of its simple extensions, can be extrapolated to energies much higher than previously thought, maybe as high as the Planck mass.
Having a premature energy cutoff prevents the possibility of making such perturbative extrapolations and of inferring properties of the very high-energy behaviour from present experimental data on the SM parameters. This problem is particularly acute in the case of Higgs inflation, where one wishes to link the shape of the inflaton potential to the known properties of the Higgs boson. For this reason we believe it useful to search for Starobinsky-like models free from any sub-Planckian violation of perturbative unitarity.

An important result of this paper is the finding of a new class of models that satisfy our general criterion for Starobinsky-like theories, but are also safe from the point of view of perturbative unitarity. These models are a special type of induced-gravity theories~\cite{induced} and, for this reason, will be called here ``induced inflation". 

\section{General criterion for single-field Starobinsky-like theories}

We consider the general Lagrangian of a single real scalar field $\phi$ coupled to gravity
\be
{\cal L}=\sqrt{-g} \left[ \frac{\Omega (\phi)}{2} R -\frac{K(\phi)}{2} (\partial \phi )^2 -U(\phi) \right] \, ,
\label{lagJ}
\ee
where we set the reduced Planck mass $\mpl =2.4\times 10^{18}$~GeV equal to one. Here the function $\Omega$ describes the non-minimal gravitational coupling, $K$ describes a possibly non-canonical kinetic term, and $U$ is the scalar potential. In order to recover Einstein gravity with (nearly) vanishing cosmological constant in the universe today,
we require that $\Omega (\langle \phi \rangle )=1$ and $U (\langle \phi \rangle )=0$ at the vacuum configuration of the field $\phi$.

We can go to a field basis where the pure gravitational term is canonical (the Einstein frame) by making the conformal transformation $g_{\mu \nu} \to \Omega^{-1}(\phi) g_{\mu\nu}$. The Lagrangian in \eq{lagJ} then becomes
\be
{\cal L}=\sqrt{-g} \left[ \frac{R}{2}  -\frac{1}{2}\left( \frac{K}{\Omega}+\frac{3{\Omega^\prime}^2}{2\Omega^2}\right) (\partial \phi )^2 -\frac{U}{\Omega^2} \right] \, .
\label{lagE}
\ee

Finally, it is convenient to redefine the field $\phi$ and render its kinetic term canonical. This is achieved with the field transformation
\be
\frac{d\chi}{d\phi}=\sqrt{ \frac{K}{\Omega}+\frac{3{\Omega^\prime}^2}{2\Omega^2}} \, ,
\label{chidef}
\ee
which gives
\be
{\cal L}=\sqrt{-g} \left[ \frac{R}{2}  -\frac{1}{2} (\partial \chi )^2 -V(\chi) \right] \, .
\label{lagC}
\ee
The scalar potential of the canonically-normalised field $\chi$ is given by
\be
V(\chi) =\frac{U[\phi(\chi)]}{\Omega^2[\phi(\chi)]}\, ,
\label{chipot}
\ee
where $\phi(\chi)$ is obtained by solving \eq{chidef}. 

Let us assume that, during inflation, the second term in the square root of \eq{chidef} dominates over the first one,
\be
\frac{{\Omega^\prime}^2}{\Omega} \gg K \, .
\label{distal}
\ee
We will come back later to the meaning of this assumption. Taking \eq{distal} as a valid approximation, \eq{chidef} can be solved analytically and we find
\be
\Omega =\exp \left( \sqrt{\frac 23}\ |\chi | \right) \, .
\label{chisol}
\ee
Moreover, we obtain
\be
\frac{dV}{d\chi}=\sqrt{\frac 23}\left( \frac{U^\prime}{\Omega \Omega^\prime}-\frac{2U}{\Omega^2} \right) \, ,
\ee
where the prime denotes derivatives with respect to $\phi$. Hence, the slow-roll parameter $\epsilon$ is given by
\be
\epsilon \equiv \frac 12 \left( \frac{1}{V}\frac{dV}{d\chi}\right)^2 = \frac 13 \left( \frac{\Omega U^\prime}{\Omega^\prime U} -2\right)^2.
\ee
During slow-roll inflation $\epsilon$ must be approximately zero and thus  $U^\prime /U \approx 2 \Omega^\prime /\Omega$. Solving this differential equation, we find that during inflation the potential $U$ and the non-minimal coupling $\Omega$ are related by
\be
U(\phi)= V_I\ \Omega^2(\phi) \left[ 1+{\rm O}(\sqrt{\epsilon})\right] \, .
\label{relstar}
\ee
The integration constant $V_I$ corresponds to the potential during inflation, as can be seen from \eq{chipot}. $V_I$ is related to the amplitude of the perturbations
\be
\delta_H^2=\frac{V_I}{150\pi^2\epsilon} \, ,
\ee
which has been measured to be $\delta_H =1.9\times 10^{-5}$~\cite{planck}. This determines the scale of inflation
\be
V_I=5.3\times 10^{-7}\epsilon ~~\Rightarrow ~~ V_I^{1/4} = \epsilon^{1/4}~6.6\times 10^{16}~{\rm GeV} \, .
\label{cobe}
\ee
Therefore, the relation between $U$ and $\Omega$ necessarily implies the presence of a small parameter. 

Let us write \eq{relstar} as
\be
U= V_I\ \Omega^2 \Delta  \, ,
\label{relstar2}
\ee
where $\Delta$ is a correction function, which we now compute.
We can determine $\Delta$ by requiring that the equation of motion during inflation are such that
\be
\Omega =1+c_1 R\, ,~~~~~U=c_2R^2\, ,
\label{infinf}
\ee
with $c_1$ and $c_2$ arbitrary constants. The form in \eq{infinf} insures that the effective theory during inflation (when the field kinetic term is negligible with respect to potential terms) reduces to $R+R^2$ gravity. Replacing \eq{infinf} into \eq{relstar2}, we find $\Delta \propto (1-\Omega^{-1})^2$.

As a result, the condition for the single-field Lagrangian to be Starobinsky-like is
\be
U= V_I\ (\Omega -1)^2   \, .
\label{relstar3}
\ee
Note that this form correctly satisfies the condition that $U$ vanishes at the vacuum (where $\Omega =1$). The potential of the canonically-normalised field is
\be
V(\chi) = V_I (1-\Omega^{-1})^2.
\ee
Under the assumption of \eq{distal}, $\Omega$ is given by \eq{chidef} and the slow-roll parameters $\epsilon$, $\eta$ and the number of e-folds are easily computed,
\be
\epsilon \equiv \frac 12 \left( \frac{1}{V}\frac{dV}{d\chi}\right)^2 = \frac 43 \frac{1}{(\Omega -1)^{2}} \, ,
\label{ee1}
\ee
\be
\eta  \equiv   \frac{1}{V}\frac{d^2V}{d\chi^2} = \frac 43 \frac{(2-\Omega )}{(\Omega -1)^{2}} \, ,
\label{ee2}
\ee
\be
N\equiv \int_{\chi_e}^{\chi} d\chi\ \frac{V}{dV/d\chi}\approx \frac 34 \left[ \Omega (\chi)-\Omega (\chi_e)\right]\, ,
\label{ee3}
\ee
where $\chi_e$ is the field at the end of inflation ($\epsilon \approx 1$).

Using the values relevant for the CMB ($N=62$), we can determine the scale of inflation from \eq{cobe}
\be
V_I=1.0\times 10^{-10} ~~\Rightarrow ~~ V_I^{1/4} = 7.6\times 10^{15}~{\rm GeV} \, .
\label{cobe2}
\ee
Moreover, since the spectral index and the tensor-to-scalar ratio are given by $n_s=1-6\epsilon +2\eta$ and $r=16\epsilon$, we obtain the Starobinsky-like prediction in \eq{star}, given that $\epsilon \approx 3/(4N^2)$ and $\eta\approx -1/N$ as shown in eqs.~(\ref{ee1})--(\ref{ee3}). 

This result is not surprising since, by construction, the theory defined by \eq{relstar2} leads to an effective $R+R^2$ gravity. This can be seen by writing
the equation of motion of the field $\phi$ derived from \eq{lagJ} in the limit of quasi-static field configuration (since, during inflation, kinetic terms are negligible with respect to potential terms). Assuming a non-minimal gravitational coupling ($\Omega^\prime \ne 0$) we find that during inflation
\be
\frac{\Omega^\prime}{2}R=U^\prime ~~~\Rightarrow ~~~\Omega =1+\frac{R}{4V_I}\, , ~~~U=\frac{R^2}{16V_I} \, , ~~~{\cal L}=\sqrt{-g} \left( \frac{R}{2}+ \frac{R^2}{16V_I}\right) \, .
\label{eqmot}
\ee
Therefore, inflation is driven by an effective $R+R^2$ gravity. As shown in ref.~\cite{riotto}, the corrections to \eq{star} coming from the kinetic terms lead to contributions to the spectral index $n_s$ of at most $10^{-3}$, hardly measurable by Planck (especially taking into account unremovable uncertainties due to the cosmological history). This explains the universality of the Starobinsky-like theories in predicting \eq{star}.

Let us come back to discuss the condition in \eq{distal}. By taking $K\sim 1$ and $\Omega/\Omega^\prime \sim \phi$, we see that \eq{distal} is roughly equivalent to $\Omega \gg \phi^2$. Our first observation is that this condition is always verified once the slow-roll condition $\Omega \gg 1$ holds, if the field $\phi$ takes only sub-planckian values during inflation. The limitation to sub-planckian field values is welcome when we want to extrapolate inflationary dynamics to low energy without encountering intermediate energy scales where there is loss of perturbative unitarity. Such scales would signal the existence of strong dynamics or new degrees of freedom, preventing calculability in the link between low-energy and high-energy theories.

The second observation is that the condition $\Omega \gg \phi^2$ implies the existence of a large coupling in $\Omega$. Indeed, take the example $\Omega \sim \xi \phi^n$, where $\xi$ is a coupling constant and $n$ an arbitrary exponent. In the linear case ($n=1$), the requirement $\Omega \gg \phi^2$, together with the slow-roll condition $\Omega \gg 1$, implies $\xi^{-1}\ll \phi \ll \xi$. A sufficiently broad field range implies strong coupling $\xi\gg 1$. For $n\ge 2$, the condition $\Omega \gg \phi^2$ implies $\xi \gg 1/\phi^{n-2}$ and hence the existence of strong coupling for sub-planckian fields (or for any field when $n=2$). The presence of a large coupling constant in $\Omega$ raises an issue with perturbative unitarity that will be discussed in the following sections.
   
If the condition (\ref{distal}) is not verified, then $\Omega$ is proportional to a small coupling and inflation occurs for super-planckian field values. In this case, we are departing from the regime of Starobinsky-like theories and entering the regime of chaotic inflation.

\section{$R+R^2$ inflation}

The theory is based on the Lagrangian~\cite{r2inflation}
\be
{\cal L}=\sqrt{-g} \left( \frac{R}{2} +\xi^2R^2 \right) \, .
\label{lagS}
\ee
We can adopt a dual description of the $R+R^2$ Lagrangian in terms of an auxiliary field $\phi$, whose Lagrangian is given by \eq{lagJ} with
\be
K=0\, ,~~~~\Omega =1+4\xi \phi \, ,~~~~U=\phi^2 \, .
\ee
Indeed, by integrating out the field $\phi$ using its equation of motion $\phi = \xi R$, we immediately recover \eq{lagS}. Once the theory is expressed in the Einstein frame, see \eq{lagE}, the auxiliary field $\phi$ acquires a kinetic term.

In this case, since $K=0$, \eq{chisol} is exactly valid and the scalar potential of the canonically-normalised field $\chi$ in \eq{chipot} is given by
\be
V(\chi)=V_I \left( 1-e^{-\sqrt {\frac 23} |\chi |}\right)^2\, ,~~~~V_I=\frac{1}{16\xi^2} \, .
\label{vrr}
\ee
The large value of $\xi$ necessary to reproduce the value of $V_I$ in \eq{cobe2} does not lead to any precocious violation of unitarity, since the Lagrangian for the field $\chi$ does not contain any positive power of $\xi$, see \eq{vrr}. Therefore, the $R+R^2$ theory preserves unitarity up to $\mpl$.

\section{Universal attractor inflation}
The class of universal attractor models, introduced in ref.~\cite{attractors}, are described by the Lagrangian in \eq{lagJ} with
\be
K=1\, ,~~~~\Omega =1+\xi f(\phi) \, ,~~~~U=\lambda f^2(\phi) \, ,
\label{attar}
\ee
where $f$ is an arbitrary function that vanishes on the present vacuum, taken to be $\langle \phi \rangle =0$ ({\it i.e.} $f(0)=0$).

Higgs inflation can be viewed as a special case of this class of models, once we work in the unitary gauge (eliminating the would-be Goldstone bosons from the Higgs doublets) and choose $f(\phi)=\phi^2$, where $\phi$ is interpreted as the real scalar field associated with the Higgs boson. Another particular example is chaotic inflation with quartic potential and non-minimal gravitational coupling (also called S-inflation), which corresponds again to $f(\phi)=\phi^2$.

The models defined in \eq{attar} satisfy the condition to be Starobinsky-like given in \eq{relstar3}.
Unlike the case of $R+R^2$ gravity, \eq{chisol} is not exactly valid, but holds during inflation under the assumption of \eq{distal}.
  Thus, during the inflationary phase, we can write the potential for the canonically-normalised field $\chi$ in \eq{chipot} as
\be
V(\chi)=V_I \left( 1-e^{-\sqrt {\frac 23} |\chi |}\right)^{2}\, ,~~~~V_I=\frac{\lambda}{\xi^2} \, .
\label{vrr2}
\ee
The CMB constraint in \eq{cobe2} implies the existence of a large coupling 
$\xi/\sqrt{\lambda}=10^5$. This large coupling gives a potential threat for perturbative unitarity. 

To investigate the issue, we need to expand the potential $V(\chi)$ around small field values. For universal attractor models, the potential $V(\chi)$ in \eq{chipot} becomes
\be
V(\chi ) =\lambda \left( \frac{f}{1+\xi f}\right)^2\, ,
\label{potpot}
\ee
where $f=f[\phi(\chi)]$.

For concreteness, let us take the case $f(\phi)=\phi^n$, in which \eq{chidef} becomes
\be
\frac{d\chi}{d\phi}=\sqrt{\frac{1}{(1+\xi \phi^n)}+\frac{3n^2\xi^2\phi^{2(n-1)}}{2(1+\xi \phi^n)^2} }\, .
\label{chidef2}
\ee
For $n>1$, the solution is
\be
\phi (\chi)= \chi \left[ 1+ F_1(\xi \chi^n, \xi^2 \chi^{2(n-1)})\right] ~~~~~({\rm for}~n>1)\, ,
\label{solsol}
\ee
where $F_1$ is a power series in its two arguments such that $F_1(0,0)=0$. Replacing \eq{solsol} in \eq{potpot}, we find
\be
V(\chi) = \lambda \chi^{2n} \left[ 1+ F_2(\xi \chi^n, \xi^2 \chi^{2(n-1)})\right] ~~~~~({\rm for}~n>1)\, ,
\label{solsol2}
\ee
where $F_2$ is another power series which vanishes at zero field value. Equation~(\ref{solsol2}) shows that the potential contains interactions with positive powers of the large coupling $\xi$. The most dangerous terms come from high powers of the second argument of $F_2$. They lead to a scale $\Lambda_{\rm UV}$ of violation of perturbative unitarity equal to
\be
\Lambda_{\rm UV}=\frac{\mpl}{\xi^{\frac{1}{n-1}}}.
\ee
This result is in agreement with the analysis of ref.~\cite{riotto}.

The case $n=1$ is special. In this case, since $\Omega^\prime$ is field independent, the solution of \eq{chidef2} at small field is
\be
\phi (\chi)= \frac{1}{\xi} \left[ e^{\sqrt{\frac 23} |\chi |} -1 + O(\xi^{-2})\right] ~~~~~({\rm for}~n=1)\, .
\label{solsol3}
\ee
The potential $V(\chi)$ is given by
\be
V (\chi)=  \frac{\lambda}{\xi^2} \left[ 1-e^{-\sqrt{\frac 23} |\chi |}+ O(\xi^{-2})\right]^2 ~~~~~({\rm for}~n=1)\, .
\label{solsol4}
\ee
This shows that, for $n=1$ there are no interactions with positive powers of $\xi$ and therefore unitarity is preserved up to $\mpl$. In the next section we will understand the reason why the case $n=1$ is special and does not cause problems with unitarity, unlike the other universal attractor models, including Higgs inflation ($n=2$). 

\section{Induced inflation}

We want to consider a class of single-field models, which we call ``induced inflation", based on the Lagrangian in \eq{lagJ} with
\be
K=1\, ,~~~~\Omega =\xi f(\phi) \, ,~~~~U=\lambda \left[ f(\phi) -\xi^{-1} \right]^2 \, .
\label{induced}
\ee
Here $f$ is a function of $\phi$ such that $f(\langle \phi \rangle )=\xi^{-1}$, where $\langle \phi \rangle$ is the vacuum configuration. The appearance of the same parameter $\xi$ in the definitions of both $\Omega$ and $U$ is only a consequence of our choice of working in units $\mpl =1$. For induced inflation, the function $f(\phi)$ must not contain any explicit dependence on the large parameter $\xi$, such as $f(\phi)=\phi^2+\xi^{-1}$, because this would bring the model back to the case of universal attractors. 

Induced inflation models satisfy the criterion for being Starobinsky-like, \eq{relstar3}. During inflation we use the approximation of \eq{distal} and write the potential of the canonically-normalised field as
\be
V(\chi)=V_I \left( 1-e^{-\sqrt {\frac 23} |\chi |}\right)^2\, ,~~~~V_I=\frac{\lambda}{\xi^2} \, .
\label{vrr3}
\ee

Let us now investigate the issue of unitarity, considering the concrete case $f(\phi) =\phi^n$. We need to solve \eq{chidef}, which reduces to
\be
\frac{d\chi}{d\phi} =\sqrt{\frac{1}{\xi \phi^n}+\frac{3n^2}{2\phi^2}} \, ,
\label{trenta}
\ee
with the boundary condition $\chi(v)=0$, where 
\be
v\equiv \langle \phi \rangle =\xi^{-1/n} \, .
\label{vvev}
\ee
For field values around the vacuum, the solution of \eq{trenta} is
\be
\frac{\phi}{v} = \exp \left(\sqrt{\frac 23}\ \frac{|\chi |}{n}\right) + O(v^{2}) \, .
\ee
The potential of the canonically-normalised field $\chi$ then takes the form
\be
V (\chi)=  \frac{\lambda}{\xi^2}  \left[ 1- e^{-\sqrt{\frac 23}|\chi |}+ O(\xi^{-2/n})\right]^2 \, .
\label{solsol5}
\ee
This expression coincides with \eq{vrr3} for $\xi \gg 1$, which holds when \eq{distal} is satisfied. 
Note that, for $n=2$, the sub-leading terms can be expressed in a simple analytic form, and the two previous equations can be written as
\be
\phi =\frac{1}{\sqrt{\xi}} \exp \left( \frac{|\chi |}{\sqrt{6+\xi^{-1}}}\right) ~~~~~({\rm for}~n=2)\, ,
\ee
\be
V(\chi)=\frac{\lambda}{\xi^2} \left[ 1-\exp \left( - \frac {2 |\chi |}{\sqrt{6+\xi^{-1}}}\right)\right]^2~~~~~({\rm for}~n=2)\, .
\label{super2}
\ee
The unitarity cutoff for the induced inflation with $n=2$ was previously discussed in the context of supergravity inflation \cite{pallis}. 

The potential $V(\chi)$ in \eq{solsol5} contains no terms enhanced by the large coupling $\xi/\sqrt{\lambda}$, hence there are no sources of unitarity violation. The crucial difference between the behaviour of induced inflation and universal attractor models with respect to unitarity can be traced back to the different kinetic terms in the Einstein frame at the field vacuum. In induced inflation, because of the non-zero vacuum expectation value of $\phi$, 
the kinetic term has a large coefficient at the vacuum
\be
\frac{{\cal L}}{\sqrt{-g}} =  -\frac{1}{2}\left. \left( \frac{K}{\Omega}+\frac{3{\Omega^\prime}^2}{2\Omega^2}\right)\right|_{\phi=\langle \phi \rangle} (\partial \phi )^2 = -\frac 34\ n^2 \xi^{2/n} (\partial \phi )^2~~~~{\rm (induced~inflation)}\, .
\label{lagEE}
\ee
The wave-function rescaling needed to go to the canonically-normalised field basis, eliminates any positive power of $\xi$ in the interaction terms. Note that the induced kinetic term in \eq{lagEE}, proportional to $(\Omega^\prime/\Omega)^2$, dominates over $K/\Omega$ both on the vacuum and on the inflationary background.

Instead, in the case of universal attractors (for $n>1$), the kinetic term at the vacuum is already canonical
\be
\frac{{\cal L}}{\sqrt{-g}} =  -\frac{1}{2}\left. \left( \frac{K}{\Omega}+\frac{3{\Omega^\prime}^2}{2\Omega^2}\right)\right|_{\phi=\langle \phi \rangle} (\partial \phi )^2 = -\frac 12 (\partial \phi )^2~~~~{\rm (universal~attractors)}\, ,
\label{lagEEE}
\ee
and the large coupling $\xi$ leads to premature violation of perturbative unitarity. Unlike induced inflation, the kinetic term at the vacuum is dominated by the first term in \eq{lagEEE}, while the second term dominates during inflation.
Note that the pure $R+R^2$ gravity model has a large $\xi^2$ coefficient of the Einstein-frame kinetic term of the auxiliary field at the vacuum, in full analogy with the unitarity-safe models of induced inflation.  

The mass of the inflaton, derived from \eq{solsol5}, is universal and does not depend on $n$,
\be
m_\chi = \sqrt{\frac{4\lambda}{3\xi^2}}=3\times 10^{13}~{\rm GeV}\, .
\ee
The mass of $\chi$ lies within the 2-$\sigma$ range (although at the very high end) of the instability scale of the Higgs potential, derived by extrapolating the SM up to high energies~\cite{VSB}. Therefore, the field $\phi$, if coupled to the Higgs in the scalar potential, could play the role of stabilising the electroweak vacuum, along the lines proposed in ref.~\cite{sthreshold}. Also, the models of induced inflation put forward in this paper, once coupled to the Higgs field, can be used to unitarize Higgs inflation, generalising the mechanism proposed in ref.~\cite{uvhiggs}.  

In order to make more explicit the relation between induced inflation and universal attractors, let us make the field redefinition
\be
f(\phi ) =  f({\tilde\phi})+\xi^{-1}\, .
\label{trans}
\ee
With this transformation, the functions $K$, $\Omega$, and $U$ in \eq{induced} become
\be
K=\left( \frac{f^\prime({\tilde\phi})}{f'(
\phi)}\right)^2=
\left( \frac{f^\prime({\tilde\phi})}{f^\prime[f^{-1}(f({\tilde\phi})+\xi^{-1})]}\right)^2\, ,~~~~\Omega =1+\xi f({\tilde\phi}) \, ,~~~~U=\lambda f^2({\tilde\phi}) \, .
\label{trans2}
\ee
The functions $\Omega$ and $U$ take exactly the form of the universal attractors, see \eq{attar},
illustrating why universal attractors and induced inflation lead to the same inflationary prediction, at leading order. The reason is that their respective $\Omega$ and $U$ are identical and the two theories differ only in $K$. The different $K$ affects inflationary predictions only at subleading order, but play a crucial role for unitarity, distinguishing induced inflation from universal attractor models.

Note also that $K$ in \eq{trans2} is equal to 1 if, and only if, $f({\tilde\phi})={\tilde\phi}$. This can be seen more easily in the case $f({\tilde\phi})={\tilde\phi}^n$, in which
\be
K=\left[ \frac{\tilde\phi}{({\tilde\phi}^n+\xi^{-1})^{1/n}}\right]^{2(n-1)}.
\ee
Indeed, we find $K=1$ only for $n=1$. In the linear case $f(\phi)=\phi$, induced inflation coincides with a particular universal attractor. In other words, a universal attractor with $n=1$ is special because it also belongs to the class of induced inflation models.  

Finally, we verify that the same results can be obtained in the new field basis.
For $f({\tilde\phi})={\tilde\phi}^n$, 
the definition of the canonical field (\ref{chidef}) becomes
\be
\frac{d\chi}{dz}=
\sqrt{\frac{\xi^{- 2/ n}}{n^2(1+z)^{3-2/ n}}+\frac{3}{2(1+z)^2}} ~ ,~~~~~z\equiv \xi {\tilde\phi}^n \, .
\ee
The second term inside the square root induced by the non-minimal coupling dominates the kinetic term, both at the field vacuum $\langle{\tilde\phi}\rangle=0$ and for the inflationary background.
Solving the above equation at small field value, we obtain 
\be
{\tilde\phi}=\left( \frac{e^{\sqrt {\frac 23} |\chi |} -1}{\xi}\right)^{1/n} + O(\xi^{-2/n})  \, .
\ee
Consequently, the potential for the canonical field $\chi$ takes exactly the same form as \eq{solsol5}.
This result confirms that there is no violation of unitarity below the Planck scale in induced inflation.

\section{Conclusions}

In this paper, we have derived a general criterion that characterises all single-field models leading to Starobinsky-like inflation with the prediction of \eq{star}. The criterion is that the Lagrangian in \eq{lagJ} must be such that $U=V_I (\Omega -1)^2$ (with $V_I=10^{-10}$) and, during inflation, the condition ${\Omega^\prime}^2/\Omega \gg K$ must hold. The latter condition implies that the function $\Omega$ contains a large coupling.

The general criterion for Starobinsky-like inflation singles out two classes of models. The first one is described by universal attractors, Higgs inflation being a special case. It is known that these models lead to a precocious violation of perturbative unitarity, at an energy scale much lower than $\mpl$. This may not be a problem for inflation to occur, but leads to a serious limitation when we want to relate, in a calculable way, the SM with the  inflaton Lagrangian. Any reliable perturbative extrapolation of the SM up to the gravity scale is impeded by the loss of unitarity at the intermediate scale.

For this reason, we have proposed a second class of Starobinsky-like models that satisfy our general criterion, but are free from any unitarity violation up to the Planck scale. In these theories, which we call induced inflation, the inflaton is a scalar field with a mass of $3\times 10^{13}$~GeV and a large vacuum expectation value that determines $\mpl$. The difference with respect to the universal attractors lies only in the kinetic terms, but this difference is crucial for unitarity. The vacuum configuration of induced-inflation models is such that large kinetic terms suppress the interaction terms, taming the effect of a large coupling, which is nonetheless necessary for inflation.

\section*{Acknowledgments}

We thank A. Riotto for discussions. HML would like to thank CERN Theory Group for its hospitality during his visit to CERN.
The work of HML is supported in part by Basic Science Research Program through the National Research Foundation of Korea (NRF) funded by the Ministry of Education, Science and Technology (2013R1A1A2007919).

\end{document}